\documentclass[12pt]{article}
\usepackage[dvips]{graphicx}

\title{Strain effects at solid surfaces near the melting point}

\author{U.~Tartaglino$^{1,2,\ast}$,  E.~Tosatti$^{1,2,3}$\\
 $^1$ SISSA, Via Beirut 2, I--34014 Trieste, Italy \\
 $^2$ INFM Democritos National Simulation Center, Trieste \\
 $^3$ ICTP, Strada Costiera 11, I--34100 Trieste, Italy \\
 $^{\ast}$ FAX: +39-040-3787528; e-mail: tartagli@sissa.it
}

\date{September 9, 2002}

\begin{document}

\maketitle

{ \abstract
We investigate the effects of strain on a crystal surface close to 
the bulk melting temperature $T_m$, where surface melting usually sets in. 
Strain lowers the bulk melting point, 
so that at a fixed temperature below but close to $T_m$ the thickness 
of the quasi-liquid film is expected to grow with strain,
irrespective of sign. In addition, a strain-induced solid surface free
energy increase/decrease takes place, favoring/disfavoring surface
melting depending on the sign of strain relative to surface stress. 
In the latter case one can produce a strain-induced 
prewetting transition, where for increasing temperature the 
liquid film suddenly jumps from zero to a finite thickness. 
This phenomenology is illustrated by a realistic molecular dynamics
simulation of strained Al(110).
\vspace{1em}
}

\noindent
KEYWORDS: Surface thermodynamics, Surface melting, Surface stress,
 Prewetting, Aluminum, Molecular dynamics

\newpage

\section{Melting of a Strained Surface: Phenomenological Theory}

We consider the fate of the surface of a {\em strained} solid just below
the bulk melting temperature, and focus in particular on surface 
melting, that is on the possibility that a microscopically thin 
surface film could melt before the bulk. The phenomenological 
surface free energy variation (per unit area) produced by melting 
the solid surface, subject to a small bulk strain $\varepsilon$ 
(assumed parallel to the surface), to a thickness $l$ of liquid
can be written as 
\begin{equation}    \label{G}
 \Delta G = l\rho\lambda\left( 1-\frac{T}{T_m}\right)
          + \left(\gamma_{SL}+\gamma_{LV}-\gamma_{SV}\right) 
          + V(l)
          - \frac{1}{2} Y' l \varepsilon^2
          - \sigma^{(s)}\varepsilon\: .
\end{equation}

Here $\lambda$ is the enthalpy of fusion per unit mass and $\rho$ is
the liquid density;
thus the first, positive term represents an increase of free energy 
($T<T_m$) required for melting a surface unstrained solid film to
become a liquid film of thickness $l$. 
The second, $\Delta\gamma_{\infty}\equiv(\gamma_{SL}+\gamma_{LV}-\gamma_{SV})$,
is the free energy variation caused by replacing, at zero strain, 
the solid-vapor interface with the solid-liquid and the 
liquid-vapor pair of interfaces.
When $\Delta\gamma_{\infty} < 0 $ ordinary, strain-free
surface melting\cite{surfmelt} takes place, with $l > 0$ close to $T_m$.
The third term $V(l)$
represent effective interaction, usually repulsive,
between the solid-liquid and the 
liquid-vapor interfaces. Its typical behaviour for small $l$ is
$V(l)=|\Delta| e^{(-2l/\xi)}$ ($\Delta$ is a parameter
comparable with $\Delta\gamma_{\infty}$, $\xi$ is the correlation length
in the liquid \cite{pluis}),  while a crossover to the asymptotic
form $V(l)=H/l^2$ appears at larger $l$'s.
The fourth term represents the decrease of the elastic energy stored in the
strained solid, $(1/2)Y'l\varepsilon^2$, since the molten film
is free to expand or contract along the surface normal
($Y'$ is proportional and close to the Young modulus $Y$)\footnote{
The exact value of $Y'$ depends on how the strain is applied. If the surface
is strained in the $y$ direction ($\varepsilon_{yy}=\varepsilon$)
and the $x$ direction is free ($\sigma_{xx}=0$) in the absence of shear 
($\varepsilon_{xy}=0$), then $\sigma_{yy} = Y \varepsilon_{yy}$ and
 $Y'=(\rho/\rho_{\mathrm{solid}})Y$. The density ratio enters because 
$l$, the thickness of the molten film, is larger than the thickness of 
the original solid film.}. 
Finally $\sigma^{(s)}\varepsilon$ is a surface 
free energy change -- in principle affecting both solid-vapor and
solid-liquid interfaces -- caused by a generally nonzero surface 
stress $\sigma^{(s)}$ . 

Eq.(\ref{G}) can be rewritten as
\begin{equation}   \label{G2}
 \Delta G(l) =
  \rho\lambda \left(\frac{T_m^{\ast}(\varepsilon)-T}{T_m}\right) l
  - ( |\Delta\gamma_{\infty}| + \sigma^{(s)}\varepsilon)
  + V(l)\: ,
\end{equation}
where $T_m^{\ast}$, is 
defined as
\begin{equation}  \label{Teff}
 T_m^{\ast} (\varepsilon) =
   \left( 1 - \frac{Y'}{2\rho\lambda}\varepsilon^2\right) \, T_m\: .
\end{equation}
As the temperature reaches $T_m^{\ast}$, the thickness of the molten
layer diverges: $T_m^{\ast}(\varepsilon)$ is the melting temperature
of the solid under strain $\varepsilon$. Actually the strain
enhances the free energy of the solid phase without altering that of
the liquid phase. Thus the latter is favoured and $T_m^{\ast} < T_m$.
We note that $\Delta G(0)=0$ but
$\lim_{l\to 0^{+}}\Delta G(l) \equiv G(0^{+}) =
-(|\Delta\gamma_{\infty}|+\sigma^{(s)}\varepsilon) + V(0^{+})$
which does not generally vanish.
In fact equations
(\ref{G}) and (\ref{G2}) cease to hold for $l$ comparable with 
a monolayer or less.

\begin{figure}
 \includegraphics[width=\textwidth]{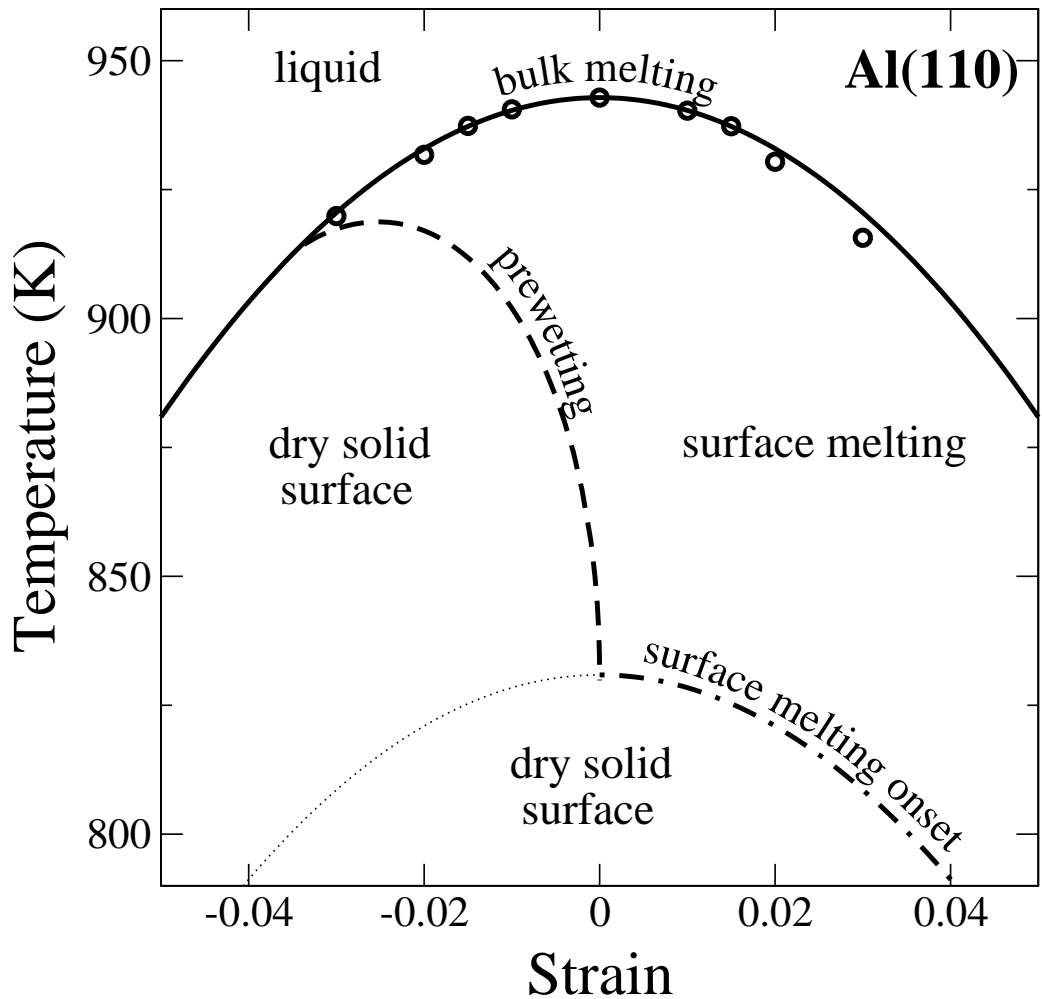}
 \caption{  \label{phasediag}
        The phase diagram of the Al(110) interface.
        The circles show the melting temperatures of the strained
        solid obtained by the molecular dynamics simulations.
        The prewetting line has been obtained by assuming
        $V(l)= |\Delta\gamma_{\infty}| e^{-2l/\xi}$ in
        equation (\ref{G}).
        Note the large strains we are considering.}
\end{figure}

The predicted strain-temperature phase diagram for realistic
parameters mimicking aluminium, and
obtained minimizing the $\Delta G(l)$ with respect to $l$ is
shown in Fig.~\ref{phasediag}.
At $T=T_m^{\ast}(\varepsilon)$ (solid line) there is
melting of the strained bulk. 
The quadratic 
decrease of $T_m^{\ast}$ with increasing strain is visible.
Surface melting appears below $T_m^{\ast}$ when, for increasing $T$,
$\Delta G(l)$ first develops a minimum for a finite value of $l$.
The nominal onset temperature
for surface melting -- the temperature where the solid surface 
is first wetted by an infinitesimal liquid film -- is attained 
at $T=T_w$ where  $d(\Delta G)/dl = 0 $ at $l=0$. This 
wetting temperature (dot-dashed line) is, similarly 
to bulk melting, quadratically depressed by strain. 

Between these two temperatures,  $T_w$ and $T=T_m^{\ast}$,
the quasi-liquid film thickness grows from zero to infinity.
The divergence at $T_m^{\ast}$ is power law  
$ l \sim (T_m^{\ast}-T)^{-1/3}$ for $H>0$,
but only logarithmic $ l \sim  |log(T_m^{\ast}-T)| $ for $H=0$.
A crossover between the two regimes appears when the derivative
of $H/l^2$ with respect to $l$ becomes comparable with that of
$|\Delta\gamma_{\infty}| e^{-2l/\xi}$:
in Al $H \sim 0.56\times 10^{-21}$ Joule\cite{chen} and
this crossover should take place at about
0.5 K below the melting point, with $l\sim 14$\,\AA.

A continuous growth of the liquid film thickness is typical of
regular, complete wetting of the solid substrate by its own melt.
For other surfaces or materials wetting may not occur at all before
bulk melting ($T_w = T_m^{\ast}$), and we have the so-called
surface non-melting\cite{nonmelt1,nonmelt2,denier}.
A discontinuous growth, with a jump in the liquid thickness,
could theoretically take place in case of a so-called prewetting
\cite{prewetting} transition.
The strain-free surfaces that have been studied so far,
experimentally as well as theoretically, were found to exhibit mostly
complete surface melting, or surface nonmelting. An intermediate
type of behavior known as incomplete surface melting, where the liquid
film thickness levels off to a finite value below $T=T_m$, 
discontinuously jumping to infinity at $T=T_m$, has also been described
\cite{incompl} and can be seen as a rather special case of prewetting.
However no case of regular prewetting, with a finite jump
of the liquid film thickness, has so far been described in 
surface melting.

As it turns out, our simple model also predicts that surfaces that
exhibit complete surface
melting could be caused to develop a prewetting transition 
by means of external strain.
That is due to the inevitable presence of nonzero surface 
stress at the solid surface, or more correctly at all interfaces
involving the solid.\cite{frenken}
The $\sigma\epsilon$ term provokes a shift,
linear to first order in the strain, 
of the interface free energy balance $\Delta\gamma_{\infty}$. 
Depending on that shifted value 
the minimum of $\Delta G(l)$ can be either
negative with respect to the crystalline surface -- thus supporting
a stable liquid film -- or positive, in which case the liquid 
film is metastable, supporting a stable dry solid surface.
In that case there is 
a whole range of strain values where $\Delta G(l)$ changes its sign
from positive to negative for increasing temperature, generating
a prewetting transition phase line, where $l$ jumps from
zero to a finite value. For even larger negative strain 
magnitudes the surface behavior is eventually predicted to 
become nonmelting.

\section{Simulations of Strained Surface Melting}

To verify the above simple theory, we simulated the thermal 
behavior of Al(110) close to the melting point, in presence 
of unidirectional in-plane strain. Molecular dynamics simulations 
of Al(110) were done in the slab geometry, both flat and bent,
\cite{passero} and the Ercolessi-Adams many body potential
for Al, whose bulk melting temperature is 943 K (to be
compared with an experimental $T_m$ of 933 K) was used. Strain was 
introduced by expanding the flat simulation box 
along the $[0\,0\,1]$ direction, or alternatively by bending the
simulated slab, while keeping it fixed in the orthogonal direction. 

\begin{figure}
 \includegraphics[width=\textwidth]{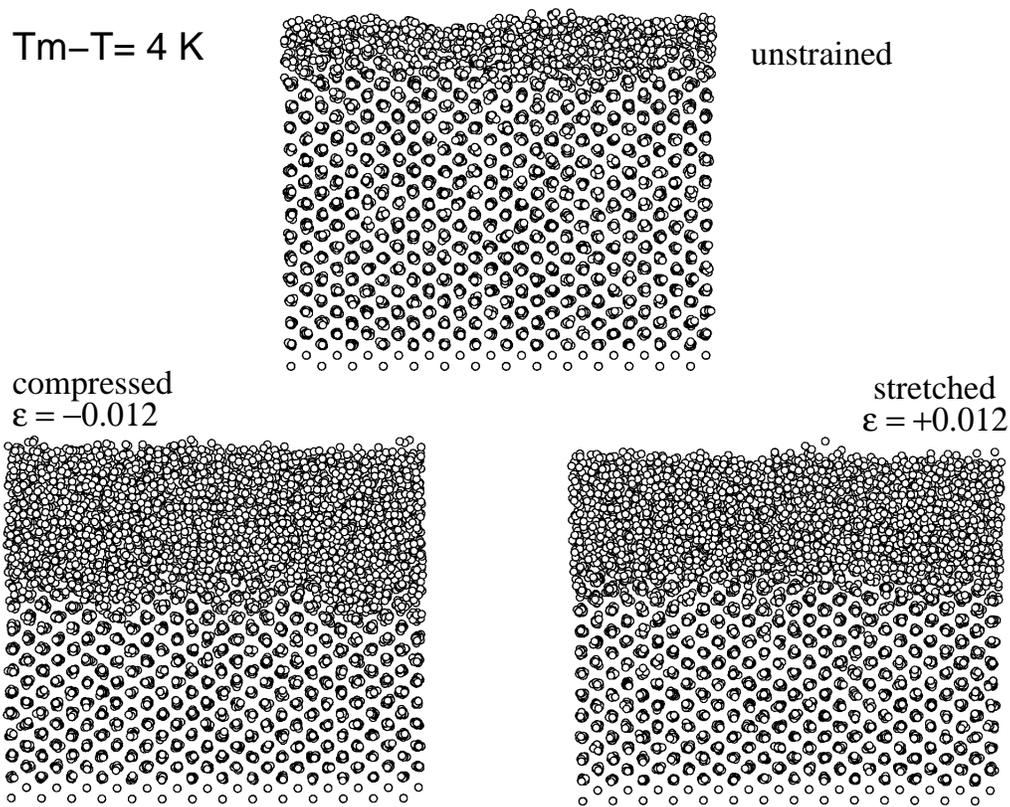}
 \caption{  \label{3casi}
           Single snapshots of the simulations of Al(110):
           side view of of three samples at the same temperature.
           Both negative and positive strain increase the molten
           film thickness, reflecting lowering of the melting
           temperature by strain.
	   Samples size: $14\times 20\times 16$ cells.
	   The atom stacking in the 20-atom rows orthogonal to the picture
	   clearly distinguishes the liquid by the solid phases.
         }
\end{figure}

The simulations, conducted with standard canonical methods
confirmed first 
of all the decrease of $T=T_m^{\ast}(\varepsilon)$ with strain. 
Fig.~\ref{3casi} shows the large increase of the melted film
occurring at fixed temperature for symmetrically positive and negative 
strains.

Subsequently, to investigate the possible presence of a prewetting 
transition, we carried out \cite{tobewritten} a series of microcanonical
simulations, 
where at each given strain the internal energy was increased stepwise.
(Fixing the energy in place of the temperature is useful
to reduce the large fluctuations of constant temperature
simulations close to a first order phase transition).
At each step the temperature as well as various structural 
correlation functions were monitored along 2.12 nanosecond long runs.
The effective melting temperature $T_m^{\ast}$ is extracted, by
determining the vertical asymptote in the plot of temperature versus molten
film thickness. Data reported in Table \ref{tm_strain} show a good
agreement with our formula (\ref{Teff}), the discrepancy at the larger
positive strains being due to anharmonic effects.
\begin{table}
\begin{center}
\begin{tabular}{| c | c| c|}
 \hline
 strain & $T_m-T_m^{\ast}\: (\pm 1 K)$ &
                $Y'/(2\rho\lambda)\: T_m \varepsilon^2$  \\
 \hline
 -0.03~ & 23.0  & 22.5 \\
 -0.02~ & 11.1  & 10.0 \\
 -0.015 &  5.5  &  5.6 \\
 -0.01~ &  2.3  &  2.5 \\
 +0.01~ &  2.5  &  2.5 \\
 +0.015 &  5.6  &  5.6 \\
 +0.02~ & 12.4  & 10.0 \\
 +0.03~ & 27.2  & 22.5 \\
 \hline
\end{tabular}
\end{center}
\caption{ \label{tm_strain}
    The decrease of the effective melting temperature with strain
The third column is obtained by formula (\ref{Teff}),
with $Y'/(2\rho\lambda) = 26.6$ for the Ercolessi-Adams potential.}
\end{table}
In this kind of simulation the possible occurrence of prewetting 
will show up, as with any other first order transition, by
hysteresis, as well as by
the occurrence of an inhomogeneous two-phase coexistence of dry
and wet portions of the surface. Because the 
surface stress of Al(110) is positive (here about 0.052\,eV/\AA$^2$
at a temperature of $900\,$K, but ab initio calculations of
Needs \cite{needs} at $T=0$ provide a value of 0.115\,eV/\AA$^2$),
it is expected that prewetting could
appear in an important way at negative strain, i.e., under
compression.

\begin{figure}
\includegraphics[width=\textwidth]{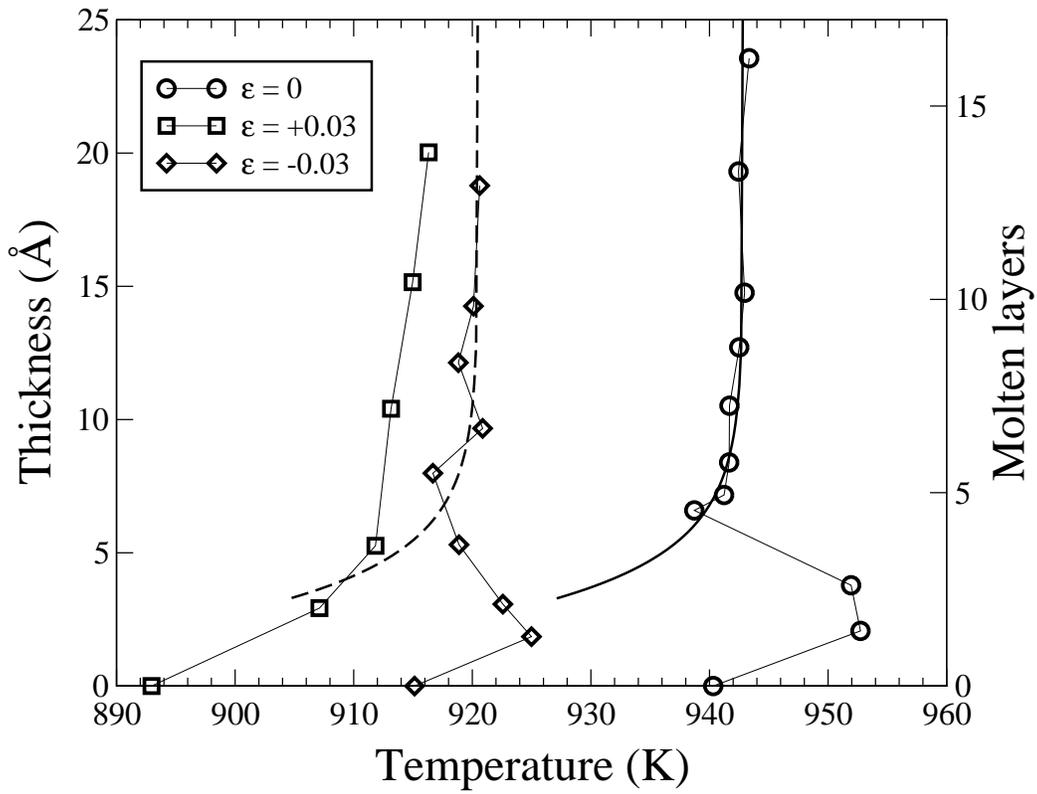}
\caption{\label{l_vs_t}Thickness of the molten layer vs Temperature.
Continuous and dashed lines are the predictions, without
considering the prewetting, for unstrained and
strain $\varepsilon=\pm 0.03$ slabs respectively.}
\end{figure}

Fig.~\ref{l_vs_t} shows the effective liquid thickness against
temperature as obtained in each simulation, for strains
of -3\%, 0, and +3\%. The expected overheating of the solid surface
is indeed observed for strains -3\%, 0, but not for +3\%, in agreement
with the snapshots. We conclude, as detailed in Ref. \cite{tobewritten},
that the Ercolessi-Adams model 
potential predicts
a prewetting transition in the surface melting of compressed 
and of strain free Al(110). The prewetting jump in the liquid layer
thickness disappears form the huge positive strain of +3\%.
The agreement with Fig.~\ref{phasediag} is qualitatively correct, although
the prewetting region turns out to be larger than expected, 
unexpectedly including zero strain.

\section{Discussion and Conclusions}

Bulk strain depresses the melting point of a solid, whose
surfaces will tend, at a given fixed temperature, to melt 
more readily than in the absence of strain. Since a tiny 
strain of order 0.0006 suffices in Al to lower $T_m$ by one hundredth
of a degree -- the typical accuracy of a surface melting
experiment\cite{accuracy} -- this suggests that random strains
could represent an important source of uncertainty in these experiments.
A combination of strained regions, resulting e.g.\ from surface
treatments, could lead to patchy behaviour close to melting, where
portions of the surface would melt and others would not.

Another consequence of strain-induced surface melting is the related
possibility to give rise to potentially interesting side effects.
If strain were introduced through a bulk longitudinal wave
for example of sufficiently low frequency $\omega$ and wavevector $k$,
one could expect a corresponding surface liquid thickness modulation of 
frequency $ 2 \omega$ and wavevector $2k$. There would also be
associated frictional damping effects worth investigating.

A newer effect of strain is the possibility to 
cause a prewetting transition in surface melting. That is 
of interest, in view of the fact that prewetting
has never so far emerged in surface melting. Although it might
prove difficult for the crystal to sustain large strains particularly
so very close to the melting point, nevertheless the entirely 
theoretical possibility of strain-induced prewetting seems 
worth addressing experimentally, as the stress needed might
in fact be quite small. In the case of Al(110) used here as
a test case, our simulations indicated prewetting already
at zero strain. While that may be an artifact of the potential
used -- existing experiments have shown no evidence of prewetting 
on Al(110)\cite{denier} -- it seems possible that prewetting could appear
with a relatively modest compression, as suggested by Fig.~\ref{phasediag}.
Prewetting is further discussed in Ref.\ \cite{tobewritten}.
\vspace{1em}

\noindent\textbf{\Large Acknowledgments}\\
This work was supported by MIUR COFIN2001, and by INFM PRA NANORUB.
Illuminating discussions with J. W. M. Frenken, D. Passerone and
F. Ercolessi are gratefully acknowledged.

\end{document}